 \documentclass[preprint2,twoside]{myaastex}

 \usepackage{amssymb}                       
 \usepackage{amsmath}                       
\usepackage[ps2pdf,bookmarks=true,breaklinks=true,hypertexnames=false,colorlinks=true,pdfstartview=FitV,linkcolor=blue,citecolor=blue,urlcolor=blue]{hyperref}
\newcommand{\myemail}{gennaro@astro.ex.ac.uk}

\newcommand{\refeqt}[1]{equation~(\ref{#1})}      

\newcommand{\refTab}[1]{Table~\ref{#1}}           

\received{year Month day}
\revised{year Month day}
\accepted{year Month day}
\ccc{code}
\cpright{none}{73 AC}

\slugcomment{\small{December 2005}}

\shorttitle{Gas Flow Across Gaps in Protoplanetary Disks}
\shortauthors{Lubow \& D'Angelo}

\begin{document}


\title{\textbf{%
Gas Flow Across Gaps in Protoplanetary Disks\footnote{%
        To appear in 
        \textsc{The Astrophysical Journal}
        sometime in 2006.
         }
      }}


\author{\textsc{Steve H. Lubow}}
\affil{Space Telescope Science Institute,
       3700 San Martin Drive, 
       Baltimore, 
       MD 21218, USA}
\email{lubow@stsci.edu\\[2mm]}
\and
\author{\vspace*{-5mm}\textsc{Gennaro D'Angelo}}
\affil{School of Physics,
       University of Exeter,
       Stocker Road,
       Exeter EX4 4QL,
       United Kingdom}
\email{\myemail\\[-5mm]}




\small

\begin{abstract}
We analyze the gas accretion flow through a planet-produced gap in a 
protoplanetary disk. We adopt the alpha disk model and ignore effects 
of planetary migration. We develop a semi-analytic, one-dimensional 
model that accounts for the effects of the planet as a mass sink and 
also carry out two-dimensional hydrodynamical simulations of a planet 
embedded in a disk.  The predictions of the mass flow rate through the 
gap based on the semi-analytic model generally agree with the 
hydrodynamical simulations at the 25\% level. Through these models, 
we are able to explore steady state disk structures and over large  
spatial ranges. The presence of an accreting $\sim1 M_J$ planet 
significantly lowers the density of the disk within a region of several 
times the planet's orbital radius. The mass flow rate across the gap 
(and onto the central star) is typically 10\% to 25\% of the mass 
accretion rate outside the orbit of the planet, for planet-to-star mass 
ratios that range from $5\times10^{-5}$ to $1\times10^{-3}$.
\end{abstract}




\section{Introduction}
\label{sec:introduction}
\noindent%
The presence of a $\sim1 M_J$  planet influences the structure of a 
circumstellar disk. The most obvious effect is the opening of a tidally 
produced gap \citep{LP93}. Recent studies of young stars such 
as CoKu Tau/4 \citep{D05}, TW Hya \citep{C02}, KH 15D \citep{H02}, HR 4796A 
\citep{S99}, and HD 141569A \citep{C03} suggest the presence of inner holes 
in circumstellar disks. One possible mechanism for producing such a hole is 
the tidal barrier created by a planet. A planet's tidal forces can, in  
principle, prevent material  from accreting across the orbit of the planet. 
The unreplenished material interior to the planet's orbit accretes onto the 
central star, thereby producing a hole. However, alternative explanations 
for apparent holes have been offered for some cases, such as dust 
segregation \citep{T01}.

This paper deals with the determination of the extent to which the  disk 
material exterior to a planet's orbit can replenish the interior material.
A  planet of $\sim 1 M_J$  opens a gap in a disk having typical parameters.
But the gap is not clean and is accompanied by accretion onto the planet 
and accretion flow across the gap \citep{AL96,B99,K99,LSA99}.
Several other past studies of planet-disk interactions have concentrated 
on analyzing torques and accretion flows onto planets. They also have 
sometimes reported finding mass flow through gaps \citep[e.g.,][]{W03}, 
while other studies have reported little inflow \citep[e.g.,][]{B03}.

The extent of the flow through the gap depends on the planet mass and disk 
properties. In the study of a $1 M_J$ planet that orbits a $1 M_{\odot}$ 
star by \citet{LSA99}, very little mass was found to accrete interior 
to the planet's orbit, not surprisingly, with nonaccreting boundary 
conditions at the inner boundary located at 0.3 times the orbital radius of 
the planet. On the other hand, if the interior disk was initially strongly 
depleted of material, then there was a substantial accretion flow that built 
up the interior disk. In this configuration, the accretion rate into the 
interior disk is comparable to the rate onto the planet. A similar density 
profile was considered recently by \citet{Q04} to model CoKu Tau/4. 
Consequently, the accretion rate onto the central star depends on the density 
distribution and the inner boundary conditions.

Even low mass planets which do not open gaps can affect the global structure 
of a circumstellar disk. The previous studies of flow across a gap are 
largely based on multidimensional hydrodynamical simulations. They are 
limited by their dynamic range in space, which typically involve a region 
that ranges from 0.4 to 6 times the orbit radius of the planet. A smaller 
inner boundary increases the computational overhead, due the need to take 
shorter timesteps because of the higher rotational velocity. Simulations 
are limited in time to at most a few thousand orbits of the planet, which 
is less than the viscous evolution timescale of a typical protostellar disk.
To overcome these limitations and provide some physical insight, we develop 
in Section~\ref{1d} a semi-analytic model for the steady state flow across 
the gap. This model depends on a key parameter, the accretion efficiency, 
which we determine by two-dimensional simulations described in 
Section~\ref{2d}. The semi-analytic model predicts the rate of accretion 
past the planet and the steady state density profile. The predictions are 
compared with results of numerical simulations. Section~\ref{sum} contains 
the conclusions.
\pagestyle{myheadings}
\markboth{\hfill Gas Flow Across Gaps in Protoplanetary Disks \hfill}%
       {\hfill \textsc{S.\ H.\ Lubow \& G. D'Angelo} \hfill}

\section[]{Semi-analytic Model \label{1d}}
\noindent%
We develop a one-dimensional, semi-analytic model for a circumstellar disk 
containing a planet. We assume that the planet lies on a fixed circular orbit 
and neglect effects of migration. In this model, we separate the gap region 
from the main disk region.
We assume that the tidal torques on the disk exerted by the planet are 
confined to the gap region. In reality, waves generated by the planet 
propagate into the main disk and exert additional torques, which we ignore.
But, shocks and other processes limit the extent of wave
propagation \citep{B02}.

We adopt the alpha disk model for the disk turbulence throughout. If the 
disk turbulence is due to magnetic fields, then the presence of a planet 
can affect the nature of the turbulence \citep{NP03,W03}. We do not account 
for such effects.

We model the effect of the planet on the main disk as a mass sink that lies 
on the planet's orbit. The mass sink description utilizes the accretion 
efficiency parameter 
\begin{equation}
\mathcal{E} = \frac{\dot{M}_p}{3 \pi \nu_p \Sigma_p},
\label{E}
\end{equation}
where $\dot{M}_p$ is the accretion rate onto the planet and $\nu_p$ is the 
turbulent kinematic viscosity. Density $\Sigma_p$ is the disk density at 
the location of the planet, based on a smooth continuation (interpolation) 
of the density profile outside the gap to the location of the planet
\citep{LSA99}. For a narrow gap, this density is the density just 
outside the gap. The denominator is recognized as the standard from for the 
steady state rate of accretion through a viscous disk, far from the disk 
inner boundary \citep{LBP74}. Since the accretion rates scale with $\nu$, 
we expect and find  that $\mathcal{E}$ is  fairly independent of $\nu$ 
(and $\Sigma_p$) for fixed $H/r$, as is consistent with the results of 
\citet{K99}. The efficiency depends mainly on the planet mass $M_p$.
One worrisome result of the numerical simulations is that $\mathcal{E}$ is 
found to be greater than unity. One might think that this suggests the 
planet is accreting more mass than the disk can supply in a steady state. 
We show later that steady state accretion flows can and usually do have
$\mathcal{E}$ greater than unity.

The density structure of a circumstellar accretion disk evolves in time 
from some initial state. The initial disk state of a protostellar disk 
is not known. Disks tend to evolve towards a steady state in which the 
accretion rate is independent of radius \citep{LBP74,P81}.
The disk evolutionary timescale in the vicinity of planets that lie 
within a 10 AU of the central star is shorter than the global evolutionary 
timescale for a 100 AU disk, and so a steady state will  be even more 
likely achieved on these smaller scales. Consequently, we seek steady 
state solutions for the disk structure.

\subsection[]{One-Dimensional Equations}
\noindent%
With the above model, the one-dimensional equations of mass conservation 
and azimuthal force balance for a disk with radial velocity $u$, angular 
velocity $\Omega$, and surface density $\Sigma$ in a steady state
($\partial \Sigma/\partial t =0,%
\partial u/\partial t =0,%
\partial v/\partial t =0$) Keplerian disk are
\begin{equation}
\frac{1}{r} \frac{d\, ( r \Sigma u)}{d r} =%
 - \frac{3\, \nu \, \Sigma\, \mathcal{E}}{2 r}%
 \delta(r - r_p)
\label{mdot}
\end{equation}
and
\begin{equation}
r^2 \Omega \Sigma u   =%
 -\frac{d\, (3 \, \mu \, \Omega\, r^2)}{d r} + 2 r \Sigma \Lambda(r),
\label{su}
\end{equation}
where $\mu = \nu \Sigma$ with kinematic turbulent viscosity $\nu$, and 
$\delta$ is the Dirac delta function. $\Lambda(r)$ is the torque density 
per unit mass produced by the tidal field of the planet 
\begin{equation}
\Lambda(r) = Sign(r-r_{p}) \frac{f q^2 G M_*}{2 r}%
 \left( \frac{r}{\Delta_{p}} \right)^4
\end{equation}
where $f$ is a constant of order unity and $\Delta_{p}$ is the maximum of 
$H$ and $|r-r_{p}|$. These equations are identical to those in \citet{LP86},
except that we have added to sink term in the equation of continuity
(\ref{mdot}).

The determination of the accretion efficiency $\mathcal{E}$ in \refeqt{mdot} 
requires a model for the capture of gas by the planet. We shall rely on 
two-dimensional simulations to determine its value. We use the 
one-dimensional model to provide initial conditions for the two-dimensional 
simulations. To compose the one-dimensional model, we divide space into two 
regions: the gap region and the main disk (nongap) region.  
We combine the solutions in the two regions of space to obtain approximate 
steady state global solutions.  In the process, we self-consistently determine 
the accretion efficiency, which links the solutions in these two regions.

\subsection[]{Gap Region \label{gr}}
\noindent%
From previous two and three-dimensional studies discussed in the Introduction, 
we know that the gap region is complicated by the presence of nonKeplerian 
flow and shocks. We do not expect that a one-dimensional model would work 
well in describing the detailed gas flow there. But, we estimate the density 
structure $\Sigma_g$ in the gap region by assuming $d/dr \gg 1/r$ (WKB 
approximation) and $1/\Delta_{p} \gg 1/r$ in \refeqt{su} to obtain
\begin{equation}
\Sigma_{g}(r)= \Sigma_{p}%
\exp{\left[- \frac{f}{9} \frac{q^2 r_p^2 \Omega_{p} }{\nu_p}\left(\frac{r_{p}}{\Delta_{p}}\right)^3\right]},
\label{G}
\end{equation}
where $\Sigma_{p}$ is the density that would occur at $r=r_p$ in the absence 
of a gap, as defined in \refeqt{E}. We generally take constant $f$ equal to 2. 
We have used initial density profiles like this in our previous work because 
they provide good initial conditions for planet-disk simulations. They undergo 
moderate changes and appear to reach a dynamically steady state in the gap 
after a few hundred planetary orbits. The achievement of a dynamically steady 
state is expedited through the development of shocks.

On longer, viscous timescales, the gap region could undergo further structural 
changes. They arise in part because of the interaction of the gap region with 
the main disk region. The coupling between these two regions occurs through 
the accretion efficiency parameter $\mathcal{E}$, which is determined in the 
gap region by the outcome of the simulations. For a consistent solution, 
we require that the value of the accretion efficiency parameter match in the
two regions, as described in the next subsection.

\subsection[]{Main Disk Region \label{md}}
\noindent%
Outside the gap region, we ignore the presence of waves which we assume damp 
most of their energy near the planet and neglect the planetary torque 
$\Lambda$.
The planetary accretion affects the large scale accretion flow as a mass sink.
We expect that main disk region is described well by the one-dimensional model.

From \refeqt{mdot}, we have that
\begin{eqnarray}
\dot{M}_{i} = - 2 \pi r u_{i} \Sigma_{i}\\
\dot{M}_{e} = - 2 \pi r u_{e} \Sigma_{e}\\
\dot{M}_{e} = \dot{M}_{i} + 3 \pi \nu_{p} \Sigma_{p} \mathcal{E},
\label{dmot}
\end{eqnarray}
where $\dot{M}_{i}$ and  $\dot{M}_{e}$ denote the mass accretion rate 
respectively interior and exterior to the orbit of the planet, and $\nu_{p}$ 
and $\Sigma_{p}$ are as used in \refeqt{E}.

We solve \refeqt{su} subject to the zero-stress boundary condition at the 
inner edge to account for the effects of a boundary layer, where the disk 
meets the central star, as described by \citet{LBP74}. To model this effect, 
one typically applies a zero density boundary condition  at the inner edge 
\citep[e.g.,][]{PVW86}, so that $\mu(r_*)=0$, where $r_*$ is the location of 
the stellar radius. The solution is given by 
\begin{equation}
3 \pi \mu_{i}(x)= \dot{M}_{i} \left(1 - \frac{x}{x_*}\right)
\label{muid}
\end{equation}
and
\begin{equation}
3 \pi \mu_{e}(x) =  \dot{M}_{o} + \frac{C}{x},
\label{mued}
\end{equation}
where $x = \sqrt{r}$, $\mu_p=\mu(x_p),$ and $\mu_{i}$ ($\mu_{e}$) is the 
value of $\mu$ interior (exterior) to the orbit of the planet, and $C$ is 
a constant of integration.

For a narrow gap, it follows that 
\begin{equation}
\mu_{i}(x_{p}) =  \mu_{e}(x_{p}),
\label{muie}
\end{equation}
since $\mu_{i}(x_{p}) $ and $\mu_{e}(x_{p}) $ are nearly equal to the 
$\mu$-values on both sides of the planet just outside the gap.
We combine equations~(\ref{dmot}), (\ref{muid}), and (\ref{mued}) to obtain
\begin{equation}
\mu_{i}(x)= \mu_p \, \frac{x_p (x-x_*)}{x (x_p-x_*)}
\label{mui}
\end{equation}
and
\begin{equation}
\mu_{e}(x) = \mu_p \left[ 1+ \left(1-\frac{x_p}{x} \right)%
 \left( \mathcal{E} + \frac{x_*}{x_p-x_*} \right) \right],
\label{mue}
\end{equation}
where $\mu_{p}$ is the value of $\mu$  smoothly extended from the main disk 
to the planet, excluding the gap, 
$\mu_p = \mu_i(x_p)= \mu_e(x_p)$.

Notice that for $\mathcal{E}=0$ (no planet) and $x_*=0$, the solution implies 
that $\mu(x)$ is constant, which is equivalent to the condition that 
$\dot{M}=3 \pi \nu \Sigma$ is independent of $r$ in a steady state, as 
discussed above.

Far outside the orbit of the planet  at some $x_{o} \gg x_p$, the disk is 
hardly aware of the planet. We have from equations~(\ref{su}) and (\ref{mue}) 
that
\begin{equation}
\dot{M}_e= 3 \pi \mu_o = 3 \pi \mu_p \left(\mathcal{E} +%
  \frac{ x_p}{x_p-x_*}\right).
\label{me}
\end{equation}

Exterior accretion rate $\dot{M}_e$ is independent of $x$ for all $x>x_p$.
The  accretion rate onto the planet is given by
\begin{equation}
\dot{M}_p=  3 \pi \mu_p \mathcal{E},
\label{mp}
\end{equation}
and the accretion rate interior to the planet's orbit (and onto the star)
is then
\begin{equation}
\dot{M}_i=  \dot{M}_e-\dot{M}_p= \frac{ 3 \pi \, \mu_p \,x_p}{x_p-x_*},
\label{mi}
\end{equation}
which is constant for $x < x_p$.

The mass flow rate ratio of accretion past the planet to accretion onto the 
planet is then
\begin{equation}
\frac{\dot{M}_i}{\dot{M}_p} = \frac{1}{(1-\sqrt{r_*/r_p}) \mathcal{E}} \, \, .
\label{rat}
\end{equation}
The ratio of the accretion rate interior to the orbit of the planet to the 
rate exterior is given by
\begin{equation}
\frac{\dot{M}_i}{\dot{M}_e} = \frac{1}{1+(1-\sqrt{r_*/r_p}) \mathcal{E}} \, \, .
\label{ratie}
\end{equation}
Ratios (\ref{rat}) and (\ref{ratie}) are independent of $\mu_p$.

The star and planet compete for accretion flow. The above accretion rate 
ratios indicate that the closer the stellar surface comes to the planet, 
the more the star accretes and the less the planet accretes. This prediction 
is confirmed in the numerical simulations discussed later (model b versus 
model g). This result is a consequence of the influence of the star in 
diverting flow from the planet. For a given kinematic viscosity $\nu(r)$ and 
accretion efficiency $\mathcal{E}$, we can determine the disk density profile 
outside the gap $\Sigma(r)=\mu(r)/\nu(r)$ from equations~(\ref{mui}) and 
(\ref{mue}). The accretion efficiency depends on the flow details within 
the gap, which we determine in Section \ref{2d} based on two-dimensional 
hydrodynamical simulations.  Equation~(\ref{rat}) determines the ratio of 
the accretion rate onto the central star to the accretion rate onto the 
planet. This ratio can be compared against results of numerical simulations.

For high mass planets (typically several Jupiter masses), the analytic model
breaks down because gravitational torques cannot be ignored in the main
disk (outside the gap). This breakdown can be seen in \refeqt{ratie}.  
For a high planet mass, the gap is very clean and the accretion efficiency 
$\mathcal{E}$ is very small. In the limit of a (nonmigrating) high planet 
mass, the disk will evolve towards a decretion disk, rather than an accretion 
disk \citep{P81}. A decretion disk has a very different description than 
the accretion disk description used here. In the high planet mass limit, 
\refeqt{ratie} predicts that mass accretes uninhibitedly past the planet, 
which is incorrect. Instead, \refeqt{ratie} applies to systems containing
lower mass planets. In that case, $\mathcal{E}$ is small when the planet mass
is small and \refeqt{ratie} provides a proper description.

Notice that there is no problem with having $\mathcal{E}$ greater than unity 
in a steady state. The accretion rate of material outside the orbit of 
the planet, $\dot{M}_e$, is greater than $3 \pi \mu_p$. The dynamic 
viscosity $\mu_p$ and typically the surface density near the planet 
(and outside the gap) are reduced due to the presence of the planet, 
in accordance with \refeqt{me}.

\section[]{Numerical Hydrodynamical Simulations \label{2d}}
\subsection[]{Description of Code \label{code}}
\noindent%
We carried out a series of numerical disk simulations in two dimensions 
using the code described in \citet{GD02,GD03}. The code calculates the 
time-evolution Navier-Stokes equations for a gaseous disk that or bits 
a central star. The disk is subject to gravitational forces from a planet 
that lies on a fixed circular orbit. The code allows for high resolution 
near the planet by using set of nested grids. We adopt a locally isothermal 
equation of state with sound speed equal to the disk aspect ratio, $H/r$, 
times the local Keplerian velocity. 
We used a constant disk aspect ratio $H/r=0.05$ throughout, implying that 
the temperature distribution scales as the inverse of the distance from the 
disk axis. Unless otherwise stated, the kinematic viscosity at the radius 
of the planet is $\nu_p=10^{-5} r_p^2 \Omega_p$, which is equivalent to 
Shakura \& Sunyaev parameter $\alpha=4\times10^{-3}$ at the location of the 
planet. We consider a range of planet masses, having a mass ratio with the 
central star of $q=5\times10^{-5}$ to $2\times10^{-3}$. Below about 
$q=4\times10^{-4}$ with a $1 M_\odot$ star, the size of the planet's 
Hill sphere is smaller than the  disk thickness and three-dimensional 
effects can be important \citep{B03}. Due to computational resource 
limitations, the simulations were carried out in two dimensions. In low end 
of the planet mass range we consider, numerical estimates of the accretion 
rate in two and three dimensions differ by only 30\% \citep{GD03}. 
So we do not expect that three-dimensional effects will substantially alter 
our conclusions.

The grid size used for these calculations consisted of a three-level 
nested-grid system with the basic level having $374\times422$ grid zones in 
the radial and azimuthal direction, respectively. The first and the second 
subgrid level had $113\times103$ and $133\times123$ grid zones, respectively.
The highest resolution achieved in the gap region around the planet was 
$\Delta r=3.7\times10^{-3} r_{p}$. The resolution in the azimuthal direction 
was equal to $\Delta r/r_{p}$.

The origin of the coordinate system is located on the star and the reference 
frame rotates about the disk axis at a rate equal to the angular velocity of 
the planet. The acceleration of the coordinate system origin relative to 
center-of-mass of the star-planet system is accounted for in the gravitational 
potential of the disk. As discussed in Section~\ref{dres}, we performed a 
convergence test on one of our models which indicates that this resolution 
is adequate. Self-gravity of the disk material is ignored.

The smoothing length of the planet potential was chosen to be 0.1 times the 
planet's Hill radius.
Two and three-dimensional models with that smoothing length yield similar 
accretion rates around a Jupiter mass planet \citep{GD03}. Some recent models 
show that variations of this smoothing by a factor of about 1.4 affects the 
accretion rate by about 1\%.

Accretion boundary conditions were generally applied to the planet.  
Such boundary conditions are currently thought to be appropriate to
the range of planet masses considered in this paper because such planets 
are believed to undergo runaway gas accretion \citep[e.g.,][]{Po96}. 
The accretion was simulated by removing mass within a radius of 0.2 Hill 
radii from the planet.  We fix a removal time-scale, which is on the order 
of 0.1 orbital periods. Tests on how the accretion rate depends on these 
two parameters are given in \citet{GD02}. The tests show that 
the accretion rate does not depend sensitively on the details of the mass 
removal parameter values, provided the removal prevents mass build-up near 
the planet.
The formulation of planetary mass accretion in \citet{LSA99} and \citet{B03} 
is completely different from that adopted here. Nonetheless, they obtain 
accretion rates that are very similar to those obtained by \citep{GD02,GD03}. 
Consequently, the evidence all suggests that the accretion rate is the rate 
that a planet would accrete under runaway accretion conditions.
It is not an artifact of the mass removal process.

The inner boundary was usually set at $0.4 r_p$, while the outer boundary was 
located at $6 r_p$.  At the inner boundary, outflow boundary conditions were 
applied.  This means that if $u_r<0$ at the first active zone, this value is 
transferred to the zones off the active grid (ghost zones). If $u_r>0$ at the 
first active zone, then $u_r$ of the ghost cells is set to zero. At the outer 
boundary, inflow and outflow of material are permitted.  The outflow is
implemented as in \citet{Go96,Go97}.  During the course of the simulation, 
the values of all physical variables are free to change at the inner and outer 
boundaries. They are not fixed by the initial analytic model.
\subsection[]{Density Determination \label{dens}}
\noindent%
For a fixed choice of $\nu(r)$ and an initial guess of accretion efficiency
$\mathcal{E}$, we determine the initial density profile from 
equations~(\ref{mui}) and (\ref{mue}) with a superimposed gap profile 
(\ref{G}). This profile is given by
\begin{equation}
\Sigma(r) = \frac{\mu(r) \Sigma_g(r)}{\nu(r) \Sigma_{p}},
\label{global}
\end{equation}
where $\mu(r)$ refers to $\mu_{i}$ for $r< r_{p}$
and $\mu_{e}$ for $r>r_{p}$.

\begin{figure}
\centering%
\resizebox{1.0\linewidth}{!}{%
\includegraphics{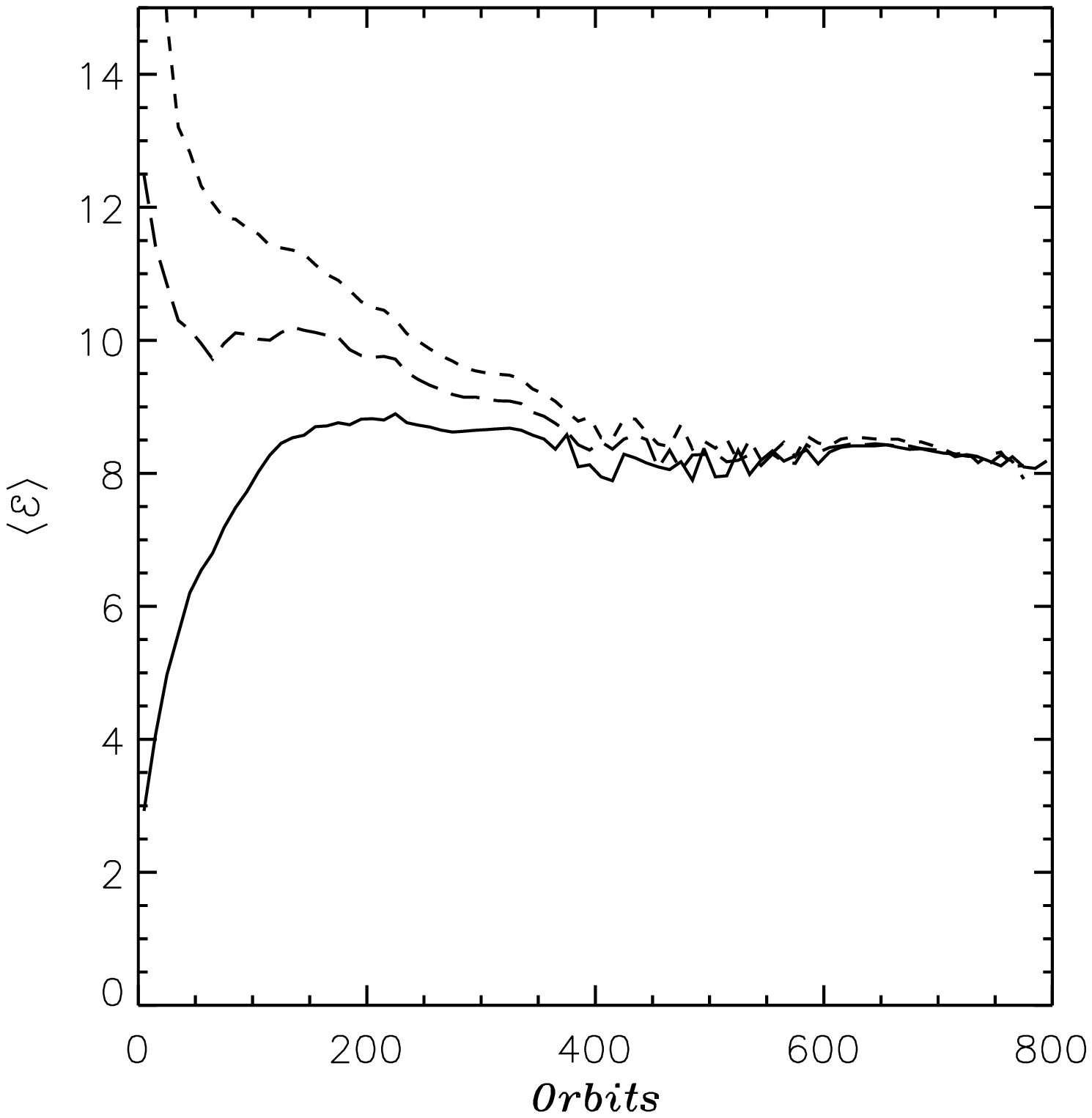}}
\caption{\small%
         Plot of the averaged accretion efficiency $\langle\mathcal{E}\rangle$, 
         averaged over the previous 10 planet orbits, versus time in planet
         orbit periods for a disk containing a $q=1 \times 10^{-3}$ planet. 
         The models in the three cases have different initial density 
         distributions in the gap region. The curves from lowest to highest
         correspond to models b, c, and d of Table 1, which have values
         of $f = 2, 1,$ and 0.5, respectively for the initial gap density
         distribution of \refeqt{G}.
         }
\label{E-fig}
\end{figure}

We generally set $r_*$ according to the location of the inner boundary.
The initial gap profile is determined by an approximate balance of viscous
torques with tidal torques, as described in Section~\ref{gr}. The value 
of $\mathcal{E}$ in the simulations is determined by the accretion rate onto 
the planet through \refeqt{mp}. We run the simulation for typically 
700-800 orbits at which point a nearly time-independent value of $\mathcal{E}$ 
is determined (see Figure~\ref{E-fig}). However, this value in general 
disagrees with the value assumed in constructing the  density profile. 
We then guess another value of $\mathcal{E}$, construct the corresponding 
density profile, and rerun the simulation. We continue iterating until the 
assumed value of $\mathcal{E}$ used for the density profile matches the 
measured value from \refeqt{mp} to within 10\%. This process typically 
requires 3 or 4 iterations.

Although we find a nearly stationary value of $\mathcal{E}$ 
(Figure~\ref{E-fig}), the determination of a strict steady state profile in 
the gap from arbitrary initial conditions would require running the code 
beyond a local viscous timescale of order $10^4$ orbits, much longer than 
700 orbits. However, the gap structure is also determined by nonlinear 
effects (shocks) that act on shorter timescales and are likely responsible 
for the near-stationarity of $\mathcal{E}$. Furthermore, from 
Figure~\ref{E-fig} we see that $\mathcal{E}$ converges in time to the same 
value, independent of initial conditions. This gives us confidence that the 
value of $\mathcal{E}$ we determine from simulations is the steady state 
value. Nonetheless, we cannot prove that the value of $\mathcal{E}$ that we 
determine is rigorously the steady state value without integrating over 
longer timescales.

\subsection[]{Description of Results \label{dres}}
\begin{table*}[t!]
 \begin{center}
 \begin{minipage}{0.90\linewidth}
 \caption{Results from simulations (see Section~\ref{dres})}
 \medskip
 \label{table}
                   \resizebox{1.0\linewidth}{!}{%
 \begin{tabular}{@{}cccccccc@{}}
  \hline
  \hline
   Model &        $q$       &    variant    & Applied & Simulated & Simulated&
   Predicted & Corrected                                                    \\
         & & & $ \mathcal{E}$   &  $ \mathcal{E}$   & $\dot{M}_i/\dot{M}_p$ &
         $\dot{M}_i/\dot{M}_p$  & $\dot{M}_i/\dot{M}_p$                     \\
  \hline
       a & $2\times10^{-3}$ &               &    3    &    2.9    &    0.52  &
                           0.91 &                                           \\
       b & $1\times10^{-3}$ &               &    8    &    8.3    &    0.38  &
                           0.34             &   0.13                        \\
       c & $1\times10^{-3}$ &  $f=1$        &    8    &    8.3    &    0.38  &
                           0.34             &   0.13                        \\
       d & $1\times10^{-3}$ & $f=0.5$       &    8    &    8.4    &    0.39  &
                           0.34             &   0.13                        \\
       e & $1\times10^{-3}$ & $2 \times\nu$ &    8    &    8.8    &    0.35  &
                           0.34             &   0.13                        \\
       f & $1\times10^{-3}$ & $\nu\propto r^{1/2}$%
                                            &    8    &    8.2    &    0.40  &
                           0.34             &   0.13                        \\
       g & $1\times10^{-3}$ & $r_{i}=0.22 r_p$%
                                            &    9    &    9.4    &    0.14  &
                           0.11             &   0.11                        \\
       h & $1\times10^{-3}$ & $H/r=0.03$    &    6     &   6.3    &    0.29  &
                           0.46             &                               \\
       i & $1\times10^{-3}$ & $\dot{M}_p=0$ &    0     &    0     &          &                                              &                               \\
       j & $5\times10^{-4}$ &               &   12     &  12.4    &    0.35  &
                           0.23             &   0.08                        \\
       k & $1\times10^{-4}$ &               &    7     &   7.3    &    0.50  &                             0.39             &   0.14                        \\
       l & $5\times10^{-5}$&                &    3     &   3.2    &     1.10 &
                          0.91              &   0.33                        \\
  \hline
 \end{tabular}
                                               }\\
 \end{minipage}
 \end{center}
\end{table*}
\noindent%
The results of the simulations are contained in Table 1. 
The default parameters consist of a kinematic viscosity 
$\nu = 10^{-5} r_p^2 \Omega_p$ which is taken to be independent of radius, 
together with the other parameters described in Section~\ref{code}.
The first column in Table 1 contains the model label. The second column 
contains the planet mass ratio. The third column contains the variant, 
if any, on the default parameters. The fourth and fifth columns contain 
respectively the accretion efficiency $\mathcal{E}$ applied to the final 
iteration and the accretion efficiency $\mathcal{E}$ obtained from the 
final simulation.
As discussed in Section~\ref{dens}, for each model we iterated on 
$\mathcal{E}$ until the applied and simulated $\mathcal{E}$ agreed 
to within 10\%. The sixth column contains the ratio of the accretion 
rate past the planet (and onto the central star) to the accretion rate 
onto the planet as obtained from the simulation. The seventh column 
contains the same quantity as the sixth, but obtained from \refeqt{rat} 
of the semi-analytic model. The eighth column contains a correction, 
described in Section~\ref{ibc}, that is applied to the values in the 
seventh column, in order to account for the location of the inner boundary.

Model b is the default $q=1\times10^{-3}$ model. The next seven models are 
variants of model b having a different initial density profiles (models c 
and d), twice the kinematic viscosity (model e), a viscosity that increases 
as $r^{1/2}$ with the same $H/r=0.05$ (model f), a smaller inner boundary 
(model g), a lower temperature  $H/r=0.03$ disk (model h), and a planet that 
does not accrete gas (model i).

A convergence test was performed on model b by using a degraded
resolution throughout the computational domain. The resolution 
was reduced by 25\%, in each direction, relative to that presented 
in Section~\ref{code}. Both in terms of accretion efficiency and
accretion rate past the planet, the outcomes agree within 2\% 
with those in the Table.
\subsection[]{Accretion Efficiency}
\noindent%
As can be seen from the Table, the accretion efficiency is largely a function 
of planet mass and is insensitive to the initial density profile (models c 
and d). It is also insensitive to the viscosity and its variation in $r$, and 
inner boundary location which are varied in models e through g, relative to 
model b. The efficiency varies from a low value of 3 for the highest and 
lowest mass planets considered, to a high value of 12 at an intermediate mass. 
This behavior is easily understood. At the high mass end, $q=2 \times 10^{-3}$ 
(model a), the tidal field of the planet is so strong that material is 
inhibited from accreting onto the planet, resulting in a lower $\mathcal{E}$ 
value relative to model b of $q=1\times10^{-3}$.

However, the accretion across the gap is somewhat less effected, since 
$\dot{M_i}/\dot{M_p}$ is  larger for $q=2\times10^{-3}$ than for 
$1\times10^{-3}$. While at the low mass end (model l), the planet's 
gravitational field has less ability to attract and accrete matter. 
For comparison, the Jupiter-mass models ($q=1\times10^{-3}$) in 
\citet{GD03} and \citet{B03} have an efficiency of about 4.
As discussed in the Introduction, a planet's accretion rate is influenced 
by the density structure outside the gap $(r>r_p)$, which in the present 
steady state case is different from that used in the above papers.

\subsection[]{Accretion Past Planet}
\noindent%
The last three columns in \refTab{table} describe the rate at which material 
flows past the orbit of the planet as a fraction of the accretion rate onto 
the planet.
The values predicted by \refeqt{rat} generally agree with the values obtained  
by the simulations at the 25\% level.

We believe model a  ($q=2 \times 10^{-3}$) is in the high mass regime where 
the semi-analytic model breaks down because gravitational torques in the main 
disk cannot be ignored, as discussed in Section~\ref{md} for decretion disks.
The disk in this case is not a true decretion disk because there is some 
accretion through the gap. However, the accretion efficiency is markedly 
lower than that in model b which has $q=1\times 10^{-3}$.
We regard model a as a  transitional disk between accretion and decretion. 
We believe this explains the discrepancy between the simulated and predicted 
accretion ratios.

For model b ($q=1\times 10^{-3}$) the simulated and predicted accretion rate 
ratios are in very good agreement, suggesting that the disk with such a planet 
behaves as an accretion disk. Agreement is also very good for model e in which
the disk viscosity has been doubled and model f, in which the $\nu$ increases 
with radius.

Model g has a smaller inner boundary located at $0.22 r_p$, instead of the 
default $0.4 r_p$, and used an initial density profile that has $r_*=0$. 
It attempts to provide more realistic coverage of the interior disk, but its 
computational domain does not cover the assumed profile extent, as do all the 
other models (which have $r_*=0.4 r_p$).  Both the simulated and predicted 
flow rates indicate that the flow rate past the planet decreases 
significantly for a smaller inner boundary radius.

Model h has a colder disk and consequently the tidal effects of the planet 
on the disk which open the gap are relatively stronger than pressure and 
viscosity. This case is intermediate between models a and b, with accretion
efficiency lower than for model b. The simulated and predicted mass flow 
ratios in columns 6 and 7 do not agree well, probably for similar reasons 
given for model a.
The flow onto the planet and across the gap are both reduced by about 25\% 
relative to model b.

Model i has nonaccreting boundary conditions onto a planet having 
$q=1\times 10^{-3}$. Consequently, we have applied an initial model having
$\mathcal{E}=0$ to the simulation. The accretion rate obtained from the 
simulation agrees with the predicated steady state accretion rate, 
\refeqt{mi}, to within 5\%. The accretion rate onto the star should be 
unaffected by the presence of the planet.

The agreement on flow rates for model j is the least satisfactory of all 
models having $H/r=0.05$ and $q \le 1\times10^{-3}$. It is not clear why 
this is the case. It may have to do with the high accretion efficiency, 
which is also observed in three-dimensional calculations \citep{GD03}. 
Fairly good agreement ($\sim 20\%$) between the simulated and 
predicted accretion rates is found for the lowest mass models k and l.
\subsection[]{Correction Due To Inner Boundary \label{ibc}}
\noindent%
For planets whose orbital radii are of order AU, the inner boundary radius 
is relatively small, $r_* \la 0.01 r_p$. It is difficult to directly simulate 
a disk with a small inner boundary, due to the shortness of timesteps required.
We estimate a correction factor for the boundary location to be applied to the
accretion rates. We saw in the previous subsection that the accretion 
efficiency $\mathcal{E}$ depends mainly on planet mass and is insensitive to 
the inner boundary location. Using the default inner boundary location of 
$0.4 r_p$ and applying \refeqt{rat}, we suggest that a correction factor to 
the flow rate ratio for a small radius inner boundary is about 
$1-\sqrt{0.4} \approx 0.37$. This factor is applied to the predicted accretion 
rate ratios (except for models a and h for which the analytic model does not 
apply, model i which does not allow accretion onto the planet, and model g 
which already has $r_*=0$). The result is in the eighth column of 
\refTab{table}.

The ratio of accretion rate onto the star to the rate onto the planet varies 
with mass ratio. The ratio is about 13\% for a $q=1\times10^{-3}$ planet and  
33\% for a $q=5\times10^{-5}$ planet.  From \refeqt{ratie} with 
$r_*=0$, the ratio of the accretion rate onto the star to the accretion rate 
outside the orbit of the planet is 11\% for a $q=1\times10^{-3}$ planet and  
25\% for a $q=5\times10^{-5}$ planet.
\subsection[]{Density Profiles}
\begin{figure*}[t!]
\centering%
\resizebox{1.00\linewidth}{!}{%
\includegraphics{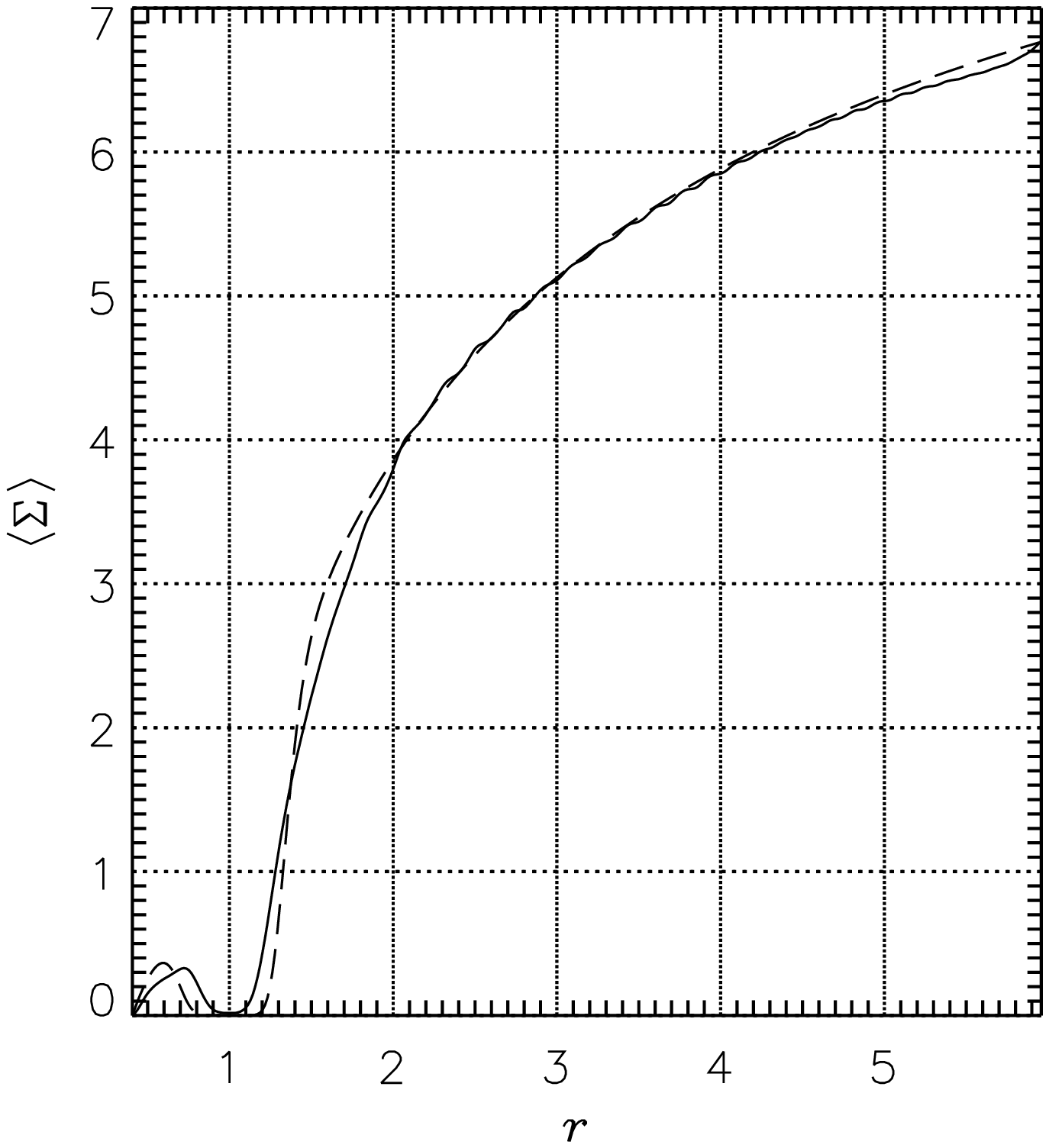}%
\includegraphics{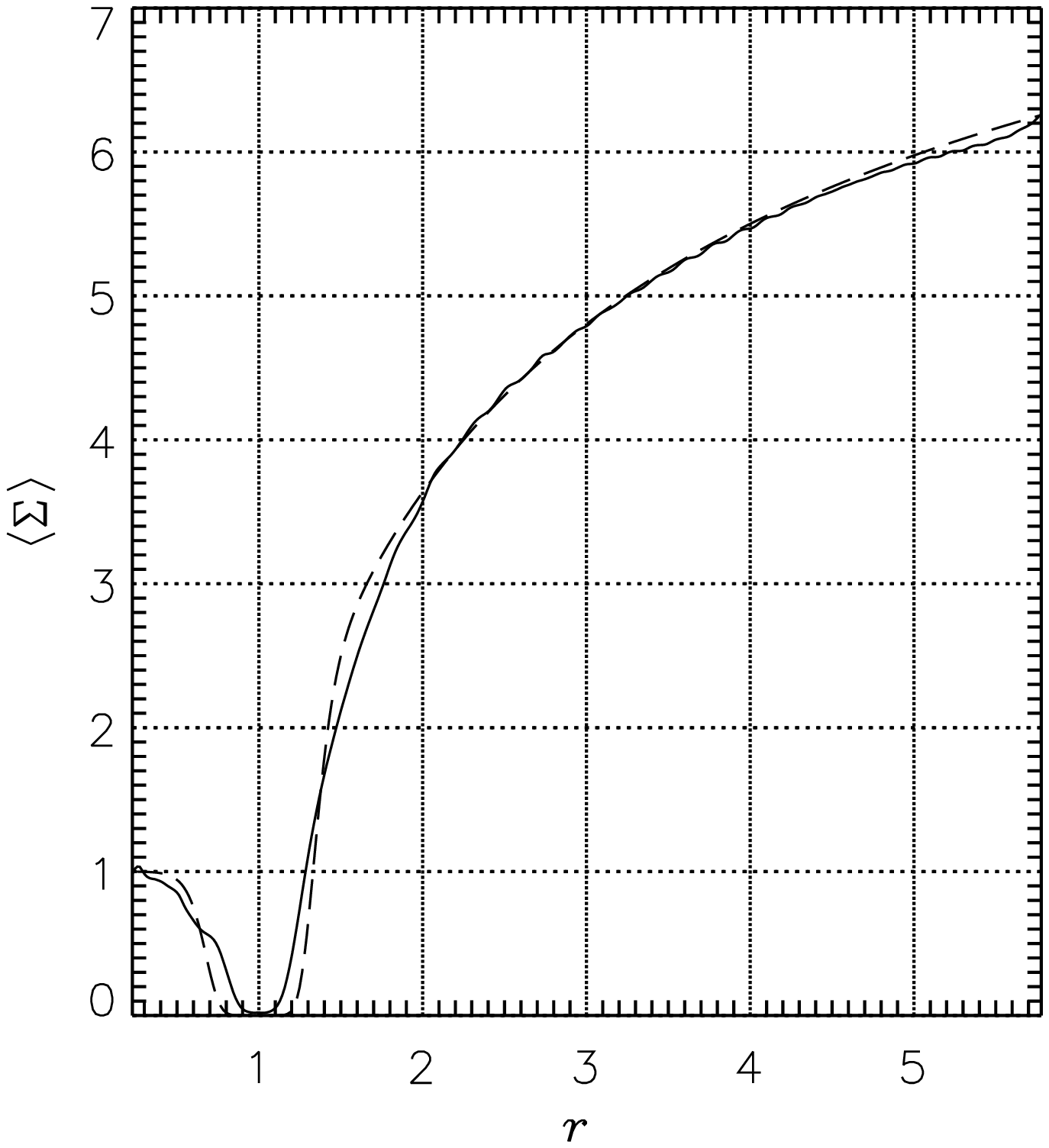}}
\resizebox{1.00\linewidth}{!}{%
\includegraphics{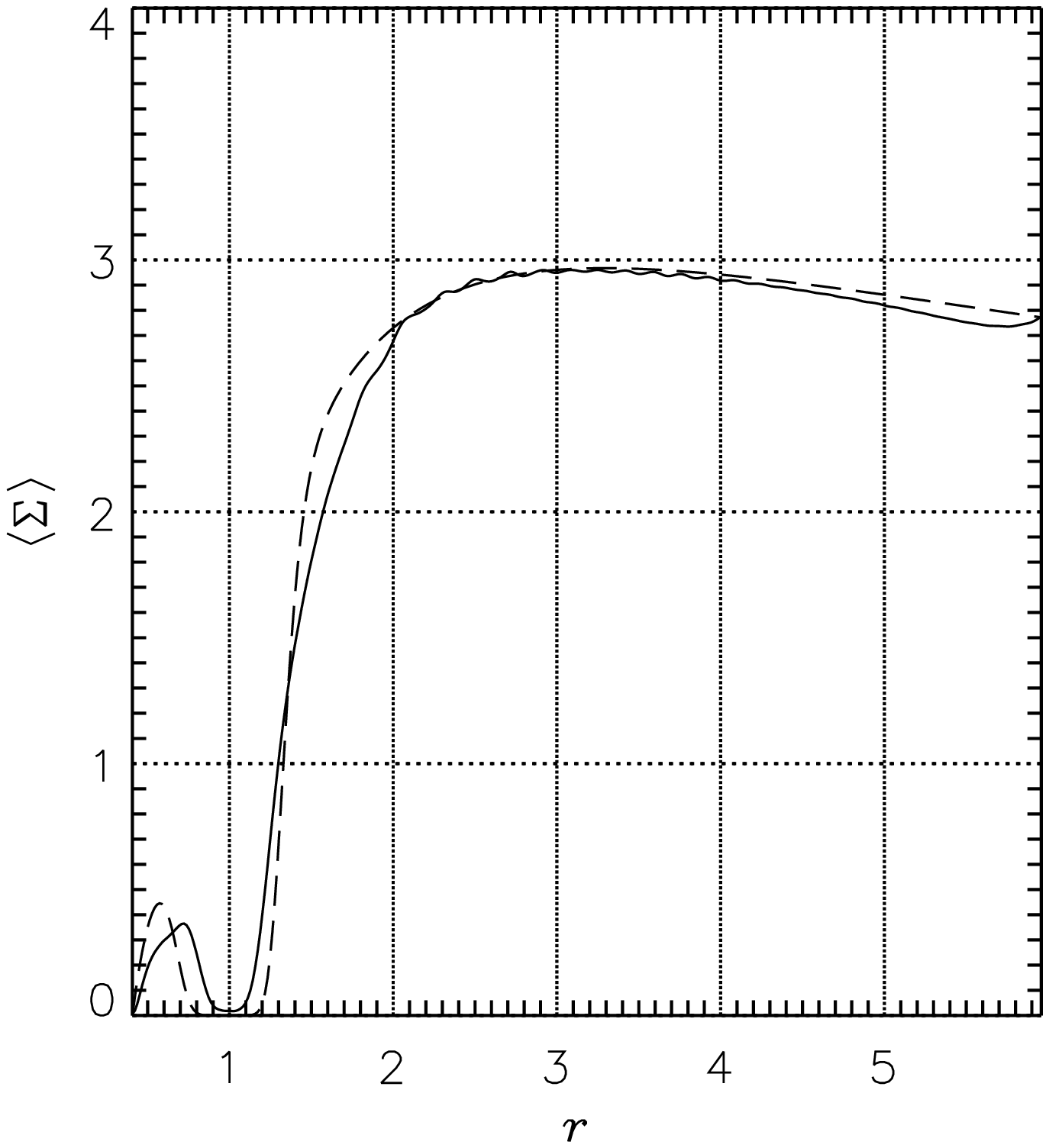}%
\includegraphics{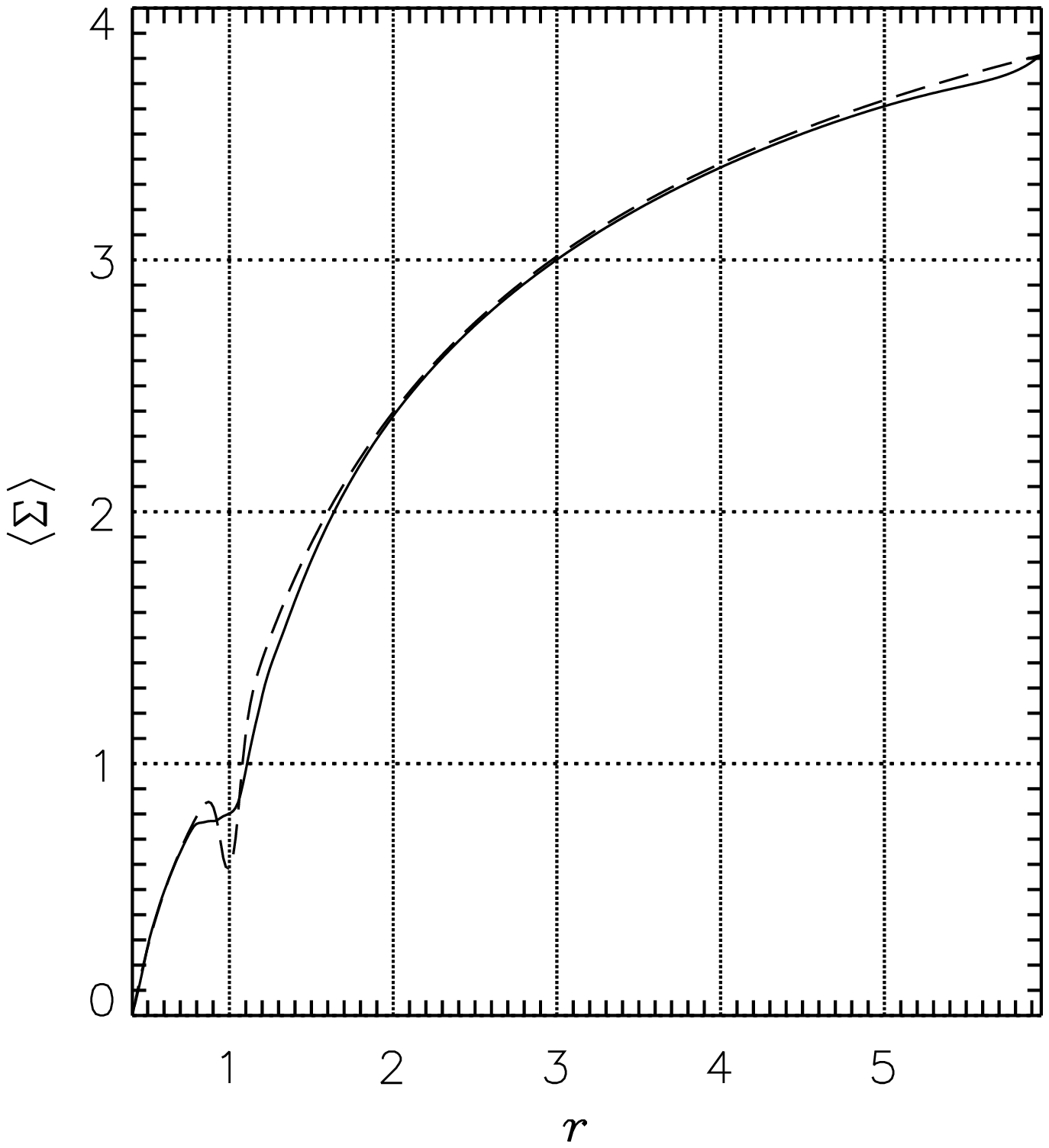}}
\caption{\small%
         Plots of azimuthally-averaged density in units of $\Sigma_p$ versus 
         radius in units of $r_p$ for the unevolved (dashed) and evolved 
         (solid) models b (top-left), g (top-right), f (bottom-left), and l
         (bottom-right) in \refTab{table}.
        }
\label{dens-evol}
\end{figure*}
\noindent%
We examine the effect of a planet on the density distribution of a steady 
state viscous disk. Figure~\ref{dens-evol} shows the assumed density profile
and evolved density profile for several models which are converged in 
$\mathcal{E}$. For all the cases, the code uses the same boundary conditions, 
as discussed in Section~\ref{code}. There are some differences in the profiles 
as a consequence of evolution, including some narrowing of the gap. Overall, 
however, the agreement between the analytic (initial) profile and the evolved 
one is very good.

\begin{figure*}
\centering%
\resizebox{1.00\linewidth}{!}{%
\includegraphics{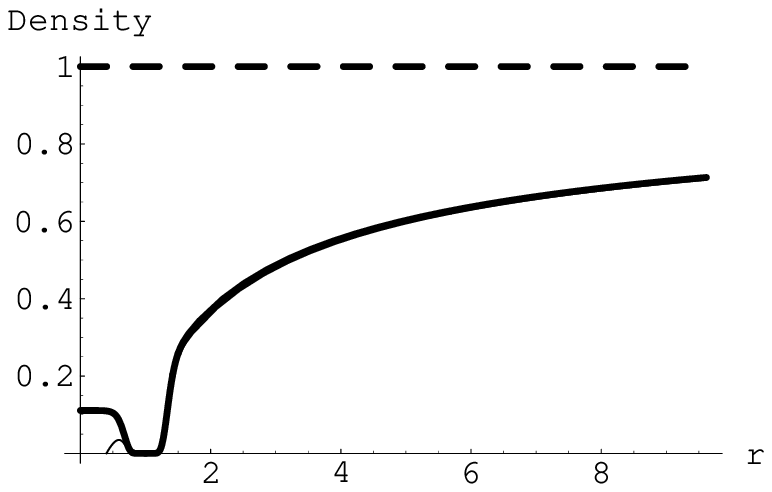}%
\includegraphics{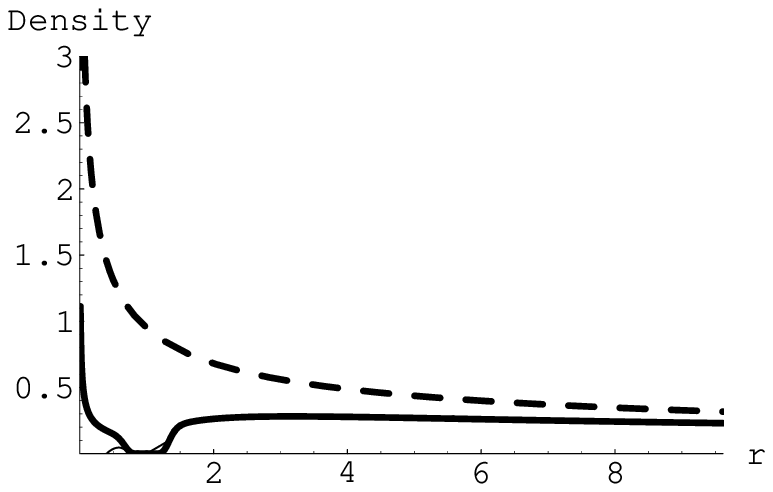}}
\caption{\small%
         Density (in units described below) for a  steady state disk 
         containing an accreting  planet, whose mass ratio is
         $q=1\times10^{-3}$, is plotted as a function of radius in units 
         of $r_p$.  The left plot (model b of Table 1) is for a constant
         disk viscosity and the right plot is for a viscosity that   
         increases in radius as $\sqrt{r}$ (model f of Table 1).  The bold 
         solid line corresponds to a disk whose inner boundary lies on a 
         star whose radius is very small compared to the orbital radius of 
         the planet ($r_* =0$). The faint solid line is for an inner 
         boundary $r_*=0.4r_p$, typically applied in multidimensional
         hydrodynamical simulations. The dashed line is for the same disk 
         without a planet. The unit of density in each plot is the density 
         of the corresponding disk without a planet at $r=r_p$.
         }
\label{dens-rb}
\end{figure*}
Global density profiles with a more realistic inner boundary location 
($r_*=0$) were constructed for a $q=1\times10^{-3}$ planet using 
$\mathcal{E}=8$  (see \refTab{table}). A global analytic density distribution 
is obtained from \refeqt{global} for some assumed viscosity function 
$\nu(r)$.  In addition, density profiles for corresponding disks without a 
planet were also determined. The results are shown in Figure~\ref{dens-rb}.
These density distributions do not contain the evolutionary effects that in 
principle could be determined by numerical simulations. But as can be seen 
from Figure 2, these effects are modest, at least over 700-800 orbits. 
Furthermore, it would be impossible to carry out numerical simulations
with a realistically small inner boundary due to the shortness of timesteps 
required.
The cases without planets correspond to steady disks that have the same 
viscosity distribution at all $r$ and the same density distribution at 
$r \gg r_p$ as the corresponding cases with planets. (For the left plot in 
Figure~\ref{dens-rb}, the density match occurs at distances larger than 
shown).
If an accreting planet were placed in a disk having the non-planet density 
distribution, indicated by dashed curves in Figure~\ref{dens-rb}, then over 
time it would evolve towards the density distributions, indicated by the 
solid curves in Figure~\ref{dens-rb}.

The accreting planet effects the disk density distribution over several 
$r_p$. For  $r \gg r_p$, the fractional density difference between the 
nonplanet and planet cases varies slowly as $\sim \sqrt{r_p/r}$. 
Consequently,  the planet's effects on the disk extend over several times 
$r_p$. Also plotted are the cases with an inner boundary of $r_*=0.4 r_p$, 
as was generally adopted in the numerical simulations. Notice that such 
cases agree well with the $r_*=0$ cases for $r > r_p$, but poorly represent 
the density at small $r$.
\section[]{Discussion and Summary \label{sum}}
\noindent%
We have investigated two-dimensional steady state configurations of a 
protostellar disk containing a planet that undergoes mass accretion, using 
both an analytic model and numerical simulations. The planet is assumed to be 
of fixed mass and on a fixed circular orbit. Planet migration and mass gain 
can be neglected at late stages of evolution where the disk mass is less than 
the planet mass. The results also suggest the outcome for the more general 
case involving mass gain and migration.

We speculate on the nonsteady effects of planet mass gain and migration. 
The mass doubling timescale for Jupiter  undergoing runaway gas accretion 
from the minimum mass solar nebula is of order the local viscous timescale 
in our model ($\sim 10^5$ years).
Consequently, the nonsteady effects of planet mass gain are of possible 
importance in that case.
Consider the case of an initially steady accretion disk in which a planet 
of mass ratio $q \la 1\times 10^{-3}$ grows and increases its mass on a 
timescale shorter than the viscous timescale. We expect that the accretion 
rate onto the central star corresponds to the steady state rate at an earlier 
time when the planet mass was lower. Therefore, we expect that the accretion 
rate onto the star to be at least $\sim$10\% of the accretion flow rate 
outside the planet's orbit in this mass range.

For an inwardly migrating planet, the available time to produce an inner hole 
is limited by  the planet's migration timescale. Furthermore, a planet that 
moves with the accretion flow (Type II migration) would not be expected to 
have much effect on the accretion rate onto the central star.
Current migration models  in the planet mass range considered here 
(0.05 to 1 $M_J$), suggest that migration occurs on about the local viscous 
timescale \citep[e.g.,][]{B03}. We then expect that an inwardly migrating 
planet  at some orbit radius would have a more substantial interior disk and  
higher accretion rate onto the star than a planet on a fixed
orbit at the same radius. Both the effects of planet mass gain and migration 
appear to lead to less depleted inner disks than the steady state models 
suggest.

Present-day multi-dimensional hydrodynamical simulations are incapable of 
following a typical protostellar disk to a steady state with spatial coverage 
down to the central star. To circumvent this problem, we have developed
a one-parameter (namely the accretion efficiency) family of semi-analytic 
steady state solutions for the main disk, excluding the gap region. These 
solutions are combined with numerical simulations that include the gap region 
to provide approximate steady state global models.

There is evidence in favor of the models being a reasonable representation of 
a steady state configuration. In particular, there is little indication of 
significant changes to the density profiles over the course of the numerical 
simulations (see Figure~\ref{dens-evol}). The density values at the inner 
boundary  in the numerical simulations (which are free to change)
change little. The steady state semi-analytic model makes a parameter-free 
prediction for the ratio of accretion onto the planet to the rate of accretion 
past the planet. The numerical simulations are in reasonable agreement with 
the predicted steady state ratio, generally within 25\%. The effects of 
changing the radius of the star on the accretion rate obtained by the 
numerical models are also in accord with the predictions of the steady state 
analytic model.

For a disk with thickness ratio $H/r=0.05$ and planet to star mass ratios $q$ 
of order $1\times 10^{-3}$ or less, about 10\% or more of the accretion flow 
continues past the orbit of the planet and onto the central star. Such 
reductions would probably still typically produce detectable accretion onto 
the central star.  Most of the accretion matter flows onto the planet.
As mentioned earlier, we have generally assumed that the planet can accrete 
the material that reaches it, in accord with the expectations of current 
planet formation theory by \citet{Po96} for the mass range we consider. 
If that assumption were incorrect to the extent that the planet does not 
accrete any mass, then the star should accrete at the rate expected in the 
absence of a $q \la 1\times 10^{-3}$ planet.

The effect of an accreting planet is to decrease the disk density over several 
times the planet's orbital radius (see Figure~\ref{dens-rb}). The presence
of an accreting planet can change the large-scale disk density distribution 
from a simple power law in radius. Inner circumstellar disks (interior to the 
orbit of the planet) occur in a steady state, even in the presence of a 
Jupiter-mass planet. The disk density inside the orbit of the planet (but 
outside the gap) is  decreased as much as about an order of magnitude, due to 
the presence of a planet. Such inner disks may be missed or underestimated 
in numerical simulations whose inner boundaries are not sufficiently close 
to the star (see Figure~\ref{dens-rb}).

Producing cleaner inner holes with lower accretion rates onto the central 
star may be possible with mass ratios in excess of $q=1\times10^{-3}$. 
For planet to star  mass ratios $q\sim5\times10^{-3}$, clean inner holes 
could result \citep{LSA99}.  But, accretion may still occur at higher planet 
masses,  if the disk becomes eccentric as a consequence of a tidal 
instability \citep{PNM01,kD06,GD04}.

For fixed disk viscosity and planet mass, a colder disk has less ability to 
penetrate the gap around a planet. The results (model h of \refTab{table}) 
show that some reduction in flow through the gap occurs with a colder disk. 
For a mass ratio $q=1\times10^{-3},$ the flow is reduced by about 25\% when
$H/r$ is reduced from 0.05 to 0.03.

In the case of the TW Hya, on the basis of low near-IR excess, there is 
evidence of dust depletion inside 4 AU, although the depletion is not 
complete \citep{C02}. Based on emission, there is evidence of 
gas accretion that is one or two orders of magnitude lower than typical 
for much younger T Tauri stars \citep{M00} . Some of the reduction 
could be due to overall gas depletion associated with its $\sim 10 Myr$  
age due to accretion. It is also possible that the presence of a planet 
of one Jupiter mass or less mass could play a role in decreasing, but 
not terminating, the accretion  in the inner region, by an order of 
magnitude. Similar considerations apply to the case of CoKu Tau/4, 
where there is evidence for dust depletion inside of 10 AU and no 
evidence of accretion \citep{D05}. We do not find that  inner disk 
accretion can be suppressed by several orders of magnitude for the 
planet mass range considered ($\la 1 M_J$). Such suppression 
may be possible with a planet of several $M_J$. But as mentioned
above, even this possibility is uncertain and requires further 
investigation.
\section*{Acknowledgments}
\noindent%
We thank Jim Pringle for helpful discussions.
We also thank the referee for useful comments that led to an improvement of 
the paper. The computations reported in this paper were performed using the 
UK Astrophysical Fluids Facility (UKAFF). GD is a UKAFF Fellow and 
acknowledges support from the STScI Visitors Program. SHL acknowledges support 
from NASA Origins of Solar Systems grant NNG04GG50G.

\begin{thebibliography}{}
\bibitem[Artymowicz \& Lubow(1996)]{AL96}
         Artymowicz, P., \& Lubow, S.~H.\ 1996, \apjl, 467, L77
\bibitem[Bate et al.(2002)]{B02}  
         Bate, M.~R., Ogilvie, G.~I., Lubow, S.~H., \& Pringle, J.~E.\ 
         2002, \mnras, 332, 575 
\bibitem[Bate et al.(2003)]{B03} 
         Bate, M.~R., Lubow, S.~H., Ogilvie, G.~I., \& Miller, K.~A.\ 
         2003, \mnras, 341, 213 
\bibitem[Bryden et al.(1999)]{B99} 
         Bryden, G., Chen, X., Lin, D.~N.~C., Nelson, R.~P., \& 
         Papaloizou, J.~C.~B.\ 
         1999, \apj, 514, 344 
\bibitem[Calvet et al.(2002)]{C02} 
         Calvet, N., D'Alessio, P., Hartmann, L., Wilner, D., Walsh, A., \& 
         Sitko, M.\ 
         2002, \apj, 568, 1008
\bibitem[Clampin et al.(2003)]{C03} 
         Clampin, M., et al.\ 2003, \aj, 126, 385 
\bibitem[D'Alessio et al.(2005)]{D05} 
         D'Alessio, P. et al.\ 2005 \apj, 621, 461
\bibitem[D'Angelo, Henning, \& Kley(2002)]{GD02} 
         D'Angelo, G., Henning, T., \& Kley, W. 2002, \aap, 385, 647
\bibitem[D'Angelo, Kley, \& Henning(2003)]{GD03} 
         D'Angelo, G., Kley, W., \& Henning, T. 2003, \apj, 586, 540
\bibitem[D'Angelo, Lubow, \& Bate(2005)]{GD04} 
         D'Angelo, G., Lubow, S., \& Bate, M.\
         2005, American Astronomical Society, DPS meeting \#37, \#31.16
\bibitem[Godon(1996)]{Go96} 
         Godon, P.\ 1996, \mnras, 282, 1107 
\bibitem[Godon(1997)]{Go97} 
         Godon, P.\ 1997, \apj, 483, 882 
\bibitem[Herbst et al.(2002)]{H02} 
         Herbst, W., et al.\ 2002, \pasp, 114, 1167 
\bibitem[Kley(1999)]{K99} 
         Kley, W.\ 1999, \mnras, 303, 696 
\bibitem[Kley \& Dirksen(2005)]{kD06} 
         Kley, W., \& Dirksen, G.\ 2005, preprint
\bibitem[Lin \& Papaloizou(1986)]{LP86} 
         Lin, D. N. C. \& Papaloizou, J. C. B. 1986, \apj, 309, 846
\bibitem[Lin \& Papaloizou(1993)]{LP93} 
         Lin, D. N. C. \& Papaloizou, J. C. B.\
         1993, in Protostars and Planets III, ed. E.H. Levy and M.S. Matthews
         (Tucson: Univ. Arizona Press), p.749
\bibitem[Lubow, Seibert, \& Artymowicz(1999)]{LSA99}
         Lubow, S.~H., Seibert, M., \& Artymowicz, P.\
         1999, \apj, 526, 1001 
\bibitem[Lynden-Bell \& Pringle(1974)]{LBP74}
         Lynden-Bell, D. \& Pringle, J.E. 1974, \mnras, 168, 603
\bibitem[Muzerolle et al.(2000)]{M00}
         Muzerolle, J., Calvet, N., Brice{\~ n}o, C., Hartmann, L., \& 
         Hillenbrand, L.\ 
         2000, \apjl, 535, L47 
\bibitem[Nelson \& Papaloizou(2003)]{NP03}
         Nelson, R.~P., \& Papaloizou, J.~C.~B.\ 2003, \mnras, 339, 993
\bibitem[Papaloizou, Nelson, \& Masset(2001)]{PNM01}
         Papaloizou, J.~C.~B., Nelson, R.~P., \& Masset, F.\ 
         2001, A\&A, 366, 263
\bibitem[Pollack et al.(1996)]{Po96} 
         Pollack, J.~B., Hubickyj, O., Bodenheimer, P., Lissauer, J.~J., 
         Podolak, M., \& Greenzweig, Y.\ 1996, Icarus, 124, 62 
\bibitem[Pringle(1981)]{P81} 
         Pringle, J.E. 1981, \araa, 19, 137
\bibitem[Pringle, Verbunt \& Wade(1986)]{PVW86}
         Pringle, J.~E., Verbunt, F., \& Wade, R.~A.\
         1986, \mnras, 221, 169
\bibitem[Pringle(1991)]{P91}
         Pringle, J.~E.\ 1991, \mnras, 248, 754 
\bibitem[Quillen et al.(2004)]{Q04} 
         Quillen, A.~C., Blackman, E.~G., Frank, A., \& Varni{\` e}re, P.\ 
         2004, \apjl, 612, L137 
\bibitem[Schneider(1999)]{S99}
         Schneider, G., et al.\ 1999, \apjl, 513, L127 
\bibitem[Takeuchi \& Artymowicz(2001)]{T01}
         Takeuchi, T., \& Artymowicz, P.\ 2001, \apj, 557, 990 
\bibitem[Winters, Balbus, \& Hawley(2003)]{W03}
         Winters, W.~F., Balbus, S.~A., \& Hawley, J.~F.\
         2003, \apj, 589, 543
\end{thebibliography}



\end{document}